\documentstyle[11pt,newpasp,twoside,epsf]{article}
\def\edcomment#1{\iffalse\marginpar{\raggedright\sl#1\/}\else\relax\fi}
\reversemarginpar

\begin{document}
\title{Using Globular Clusters to Test Newton's Law of Gravity}
\author{Riccardo Scarpa, Gianni Marconi \& Roberto Gilmozzi}

\affil{European Southern Observatory, Santiago, Chile.}

\begin{abstract}
New measurements of the velocity dispersion of the globular
cluster $\omega$ Centauri allow to trace its gravitational
potential down to an acceleration of $8\times10^{-9}$ cm s$^{-2}$.  
It is found that the
dispersion profile remains flat well inside the tidal radius as soon
as the acceleration of gravity approaches $a_0$, a result that finds its
simplest explanation within the contest of MOND. A similar behavior is
observed in the globular cluster M15 
showing this is not a peculiar features of $\omega$ Centauri.
This result is surprising and suggestive of a failure of Newton's law
at low accelerations.
\end{abstract}

\section{Introduction and Discussion}

We present the results of a study of the kinematics of the external
regions of $\omega$ Centauri and M15, a work triggered by the increasing
evidences that Newton's law of gravity may not apply to
acceleration smaller than $a_0 = 1.2\times 10^{-8}$ cm s$^{-2}$ 
(Begeman et al. 1991), as
proposed within the contest of the modified Newtonian dynamics (MOND;
Milgrom 1983). We focused on globular clusters because they are
believed to be dark matter free and hence should agree precisely with
Newton's law at any acceleration.
Interestingly, in $\omega$ Centauri  the velocity
dispersion does not show the typical Keplerian falloff remaining 
constant at large radii (Fig. 1).  It is worth to point
out that assuming a mass to light ratio M/L=1 in solar units,
appropriate for globular clusters as suggested by dynamical study
(Mandushev et al. 1991), it turns out that the dispersion
flattens for $a = 2.1\pm 0.5 \times 10^{-8}$ cm s$^{-2}$, similar to $a_0$.
A similar set of data for M15 (Drukier 1998) shows that also in this
globular cluster the velocity dispersion profile remains flat (within
errors) at large radii (Fig. 1). The flattening of the velocity
dispersion occurs for 
$a = 1.7\pm 0.6 \times 10^{-8}$ cm s$^{-2}$, again similar to $a_0$.

Though it is conceivable the observed flattening of
the velocity dispersion profile is due to either tidal heating or to a
massive dark matter halo surrounding the clusters, both scenarios
require ad hoc assumptions to make the profile flat. What is striking
is that these two clusters have very little in common, having
different masses, different positions and different orbits in the
galactic halo.  Their dynamical evolution was also different to such
an extent that while M15 is a textbook example of a globular cluster,
 $\omega$ Centauri has been claimed to be the merger of two cluster (Lee et
al. 1999) or the remnant of a dwarf galaxy (Hilker 2000).  
Thus we find the fact that the two profiles are so similar a
significant one.  Globular clusters are hundreds of times smaller and
thousands of time less massive than galaxies, nonetheless the velocity
dispersion profile of at least these two clusters precisely mimics,
both in shape and absolute acceleration, the one of elliptical
galaxies (explained invoking dark matter; Carollo et al. 1995).  There
is no reason for the flattening in globular clusters and galaxies to
occur for the same value of the acceleration. Therefore,
our result finds its simplest explanation within the framework of
MOND, supporting the suggestion that  Newton's law of gravity may fail in the
low acceleration limit.

\begin{center}
\plottwo{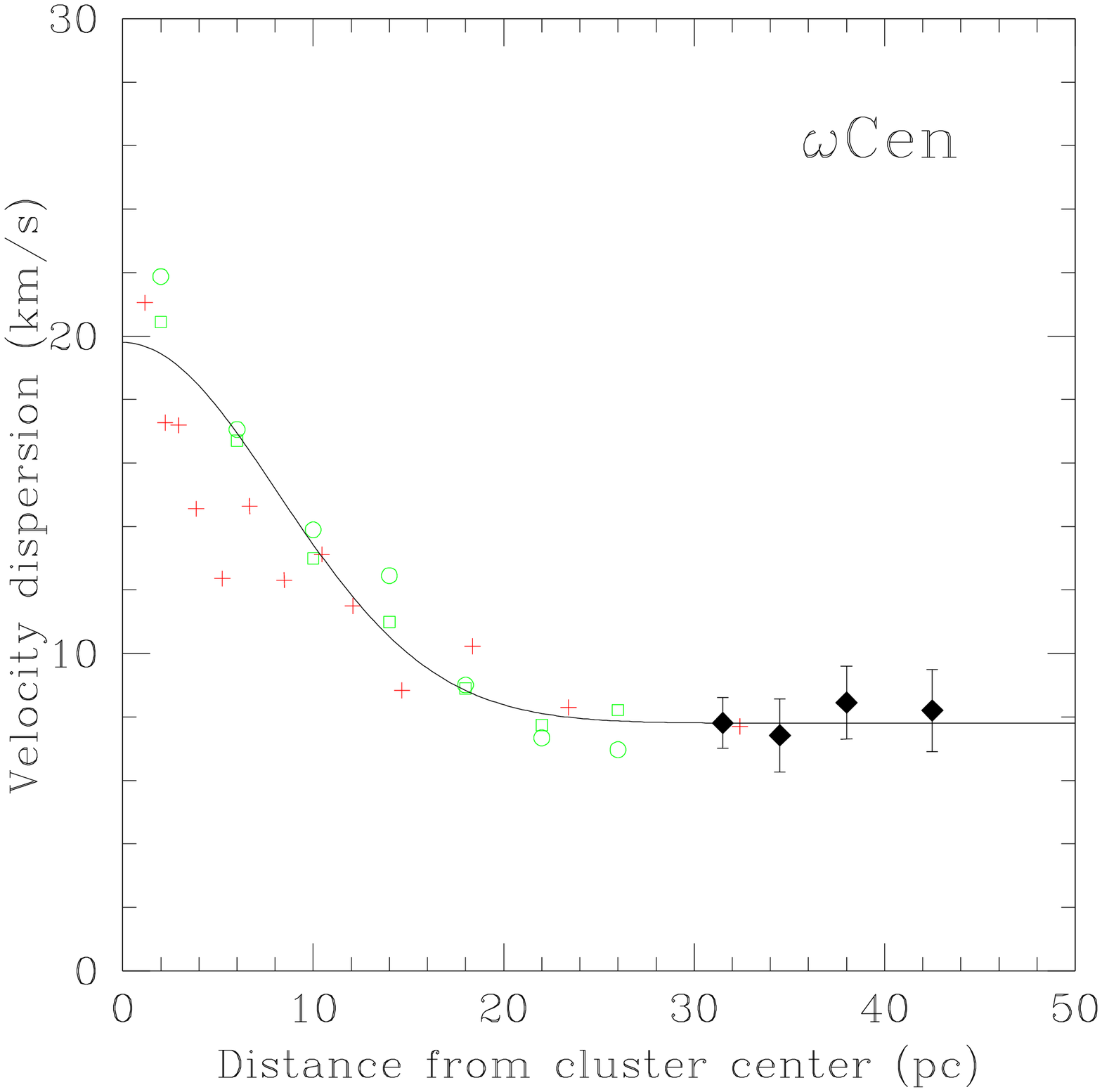}{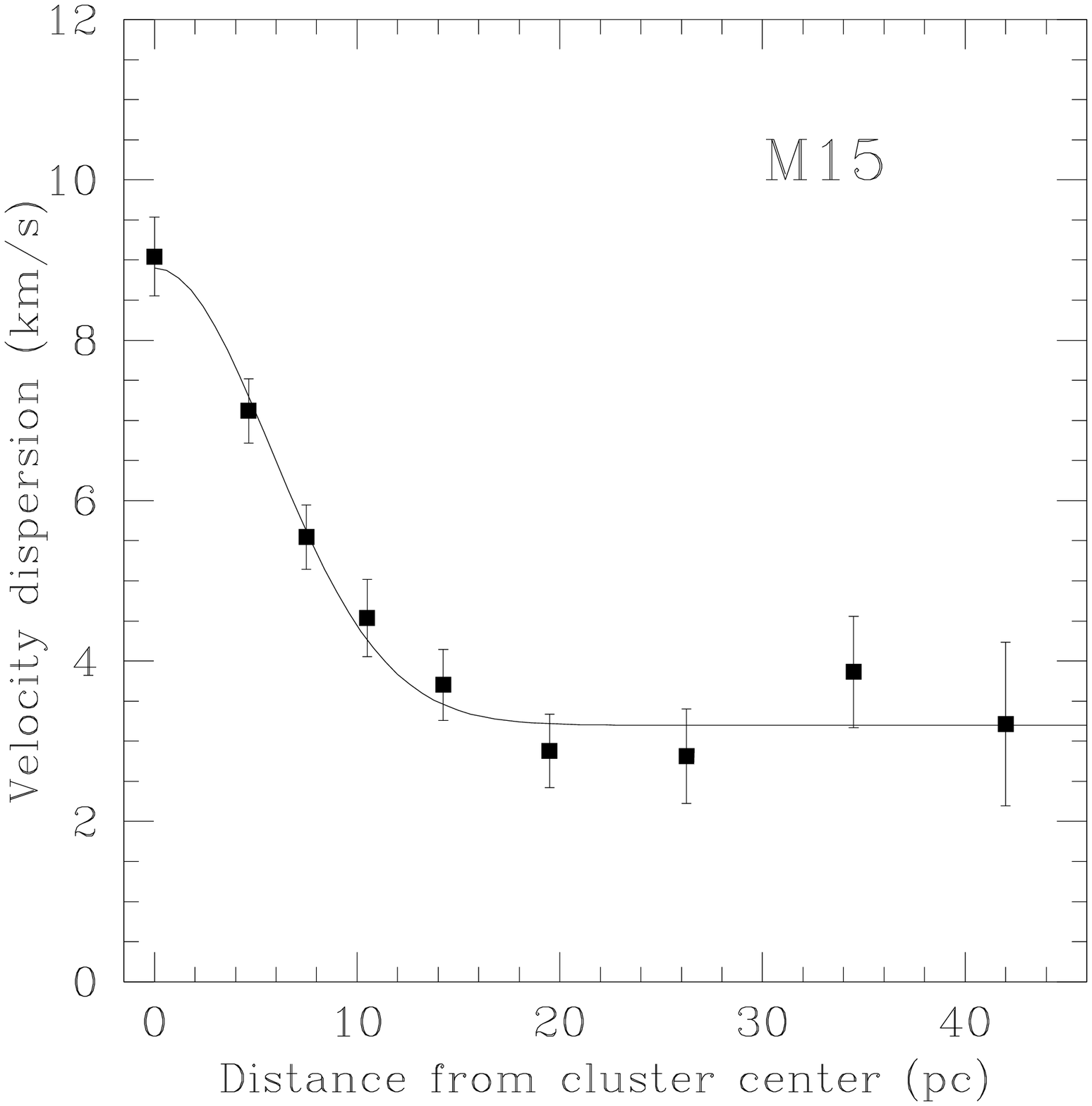}
\end{center}

\noindent
{\bf Fig 1. Left:} The velocity dispersion profile of $\omega$
Centauri. Proper motion (Circles and Squares, Van
Leeuwen et al. 2000) and  radial velocities (Crosses, Meylan et
al. 1995 and Meylan \& Mayor 1986) data  show the velocity ellipsoid 
is isotropic.  Our radial
velocities measurement (Diamonds) extend the profile to ~45
pc showing the velocity dispersion remains
constant for R$>27\pm 3$ pc.  {\bf Right:} Dispersion profile of the
globular cluster M15, from data by Drukier et al. (1998). Also in this case the 
velocity dispersion  remains basically constant for R$>18\pm 3$ pc.
In both panel the solid line is not a fit to the data. It is meant to
visualize that the velocity dispersion remains constant at large radii.


\begin{references}

\reference Begeman K.G., Broeils A.H. \& Sanders R. H. 1991, MNRAS 249, 523 
\reference Carollo C.M., de Zeeuw P.T., van der Marel R.P., et al. 1995, ApJL 441, 25 
\reference Drukier, G.A., Slavin, S.D., Cohn, H.N. et al 1998, ApJ 115, 708.
\reference Hilker M. \& Richtler T. 2000, A\&A 362, 895 
\reference Lee Y.W., Joo J.M., Sohn Y.J. et al. 1999, Nature 402, 55
\reference Mandushev G., Spassova N. \& Staneva A. 1991, A\&A 252, 94
\reference Meylan G., Mayor M., Duquennoy A. \& Dubath P. 1995, A\&A 303, 761
\reference Meylan G. \& Mayor M. 1986, A\&A 166, 122
\reference Milgrom M. 1983, ApJ 270, 365
\reference van Leeuwen F., Le Poole R. S., Reijns R.A. et al. 2000, A\&A 360, 472
\end{references}
\end{document}